# Techno-Economic Analysis of Hydrogen Production: Costs, Policies, and Scalability in the Transition to Net-Zero

**Eliseo Curcio[1,*]**

[1] Independent Researcher

*corresponding authour : *curcioeliseo@gmail.com*

*Abstract :*

This study presents a comprehensive techno-economic analysis of gray, blue, and green hydrogen production pathways, evaluating their cost structures, investment feasibility, infrastructure challenges, and policy-driven cost reductions. The findings confirm that gray hydrogen ($1.50–$2.50/kg) remains the most cost-effective today but is increasingly constrained by carbon pricing. Blue hydrogen ($2.00–$3.50/kg) offers a transitional pathway but depends on CCS costs, natural gas price volatility, and regulatory support. Green hydrogen ($3.50–$6.00/kg) is currently the most expensive but stands to benefit from declining renewable electricity costs, electrolyzer efficiency improvements, and government incentives such as the Inflation Reduction Act (IRA), which provides tax credits of up to $3.00/kg. The analysis shows that renewable electricity costs below $20–$30/MWh are essential for green hydrogen to achieve cost parity with fossil-based hydrogen. The DOE's Hydrogen Shot Initiative aims to lower green hydrogen costs to $1.00/kg by 2031, emphasizing the need for CAPEX reductions, economies of scale, and improved electrolyzer efficiency. Infrastructure remains a critical challenge, with pipeline retrofitting reducing transport costs by 50–70%, though liquefied hydrogen and chemical carriers remain costly due to energy losses and reconversion expenses. Investment trends indicate a growing shift toward green hydrogen, with over $250 billion projected by 2035, surpassing blue hydrogen's expected $100 billion. Carbon pricing above $100/ton $CO_2$ is likely to make gray hydrogen uncompetitive by 2030, accelerating the shift to low-carbon hydrogen. Hydrogen's long-term viability depends on continued cost reductions, policy incentives, and infrastructure expansion, with green hydrogen positioned as a cornerstone of the net-zero energy transition by 2035.

## I. Introduction :

The increasing urgency to decarbonize energy systems and mitigate climate change has placed hydrogen at the forefront of global energy transition strategies. As a versatile energy carrier, hydrogen has the potential to serve as a clean alternative to fossil fuels across multiple sectors, including industrial processes, transportation, and energy storage [1]. However, despite its promise, the widespread adoption of hydrogen is hindered by economic, technological, and policy challenges. The high costs associated with hydrogen production, storage, and transportation—particularly for green hydrogen—remain key barriers to scalability. Additionally, the competitiveness of hydrogen as an alternative fuel depends on evolving regulatory frameworks, investment incentives, and technological advancements [2,3]. Hydrogen production methods are broadly categorized into three pathways: gray, blue, and green, each with distinct economic and environmental trade-offs [4]. Gray hydrogen, produced from natural gas via steam methane reforming (SMR) without carbon capture, remains the dominant and lowest-cost production method today. However, its high carbon footprint is increasingly subject to regulatory pressures and carbon pricing schemes [5]. Blue hydrogen, which integrates carbon capture and storage (CCS) with SMR, offers a lower-emission alternative while leveraging existing natural gas infrastructure. Yet, its economic feasibility remains dependent on the cost and efficiency of CCS technologies and the volatility of natural gas prices [6]. Green hydrogen, produced via electrolysis using renewable electricity, presents the most sustainable solution, eliminating direct carbon emissions. However, its adoption is constrained by high electricity consumption, capital expenditures for electrolyzers, and infrastructure limitations [7].

This study aims to provide a comprehensive techno-economic analysis of hydrogen production, evaluating cost structures, financial feasibility, and the impact of policy incentives across different production pathways. A key objective is to determine the levelized cost of hydrogen (LCOH) for green, blue, and gray hydrogen under varying economic and market conditions. The study employs financial modeling tools—including net present value (NPV), internal rate of return (IRR), and sensitivity analysis—to assess investment feasibility and break-even thresholds [8,9]. In addition to production costs, the study examines hydrogen's scalability, transportation challenges, and storage requirements, which are critical factors influencing its market adoption [10]. A major component of this analysis is the evaluation of policy-driven incentives such as the U.S. Inflation Reduction Act (IRA) hydrogen production tax credits, which have been designed to enhance the economic viability of low-carbon hydrogen [11]. The study quantifies how such incentives reduce LCOH, improve investment attractiveness, and shift market preferences toward cleaner hydrogen pathways. Furthermore, the research explores how carbon pricing mechanisms, renewable energy cost reductions, and advancements in electrolyzer efficiency influence hydrogen's competitiveness over time [12].

This research is structured as follows: First, a detailed breakdown of hydrogen production costs is presented, identifying the key cost drivers for each production method. Next, transportation and storage challenges are analyzed, assessing the feasibility of hydrogen distribution via pipelines, liquefaction, and chemical carriers. The study then evaluates financial models to determine the economic feasibility of hydrogen investments under different pricing scenarios and policy conditions. Lastly, a comparative analysis of global investment trends is conducted, highlighting the shifting market dynamics as capital flows toward low-carbon hydrogen solutions [13]. By integrating financial modeling, cost assessments, policy impact analysis, and scalability considerations, this study provides a data-driven evaluation of hydrogen's economic trajectory.

## II. Methods :

### II.1. Data collection and modeling approaches :

This study employs a structured approach to evaluating the economic viability of hydrogen production, integrating financial modeling, policy analysis, and data sourced from industry reports, peer-reviewed research, and government publications [1]. The analysis focuses on quantifying the costs, investment feasibility, and policy incentives associated with green, blue, and gray hydrogen, ensuring that findings are derived from factual, data-driven insights rather than speculative assumptions [2]. The selection of data sources was guided by relevance, credibility, and applicability to hydrogen market trends. The study relies on publications from the International Journal of Hydrogen Energy, Hydrogen Council, McKinsey & Co., DNV Maritime Forecast, and legislative frameworks such as the U.S. Inflation Reduction Act (IRA) [3]. These sources provide quantifiable metrics on production costs, infrastructure challenges, and investment trends. Particular emphasis was placed on reports that outline cost breakdowns for hydrogen technologies, including capital expenditures (CAPEX), operational expenditures (OPEX), and feedstock pricing, as well as studiesevaluating the levelized cost of hydrogen (LCOH) across different production pathways [4]. The data was cross-referenced where possible to ensure consistency and remove outliers that might skew financial assessments [5]. The financial analysis applies a levelized cost of hydrogen (LCOH) framework to determine the economic feasibility of hydrogen production [6]. LCOH is calculated by dividing the total annualized cost of production by the annual hydrogen output, incorporating CAPEX, OPEX, and feedstock inputs [7]. The modeling considers electricity price variability for green hydrogen, reflecting how fluctuations in renewable energy costs impact production economics [8]. A Net Present Value (NPV) model was developed to assess investment feasibility, using a 7% weighted average cost of capital (WACC) and a 20-year project timeline [9]. The model evaluates profitability under different hydrogen pricing scenarios, identifying the break-even point at which NPV turns positive. This enables a comparative assessment of how green, blue, and gray hydrogen projects perform under various market conditions [10].

### II.2. Assumptions and scope of analysis :

Key assumptions were established to provide a consistent basis for financial projections. Electricity costs for green hydrogen were assumed to range from $20–$50/MWh, aligning with global trends in renewable energy pricing [15]. Carbon capture efficiency for blue hydrogen was modeled at 85–90%, based on industry benchmarks for CCS technology [16]. Transportation and storage costs were estimated using current infrastructure costs, acknowledging that future reductions may occur as hydrogen networks expand [17]. While demand growth is expected to play a crucial role in determining investment feasibility, the study does not account for potential market shocks or unforeseen regulatory changes that could significantly alter hydrogen's economic landscape [18]. This methodology ensures that all findings are based on factual analysis rather than hypothetical projections [19]. By integrating financial modeling, policy impact assessment, and sensitivity analysis, the study provides a comprehensive evaluation of hydrogen's commercial viability across different production pathways [20]. The financial models developed in this analysis rely on key assumptions regarding CAPEX, OPEX, electricity pricing, and policy support [22]. These assumptions serve as the foundation for evaluating the economic feasibility of hydrogen production across different technologies. Table 1 outlines the primary cost assumptions used in this study [23].

Table 1 : Key Assumptions for Hydrogen Financial Modeling

| Parameter | Green Hydrogen | Blue Hydrogen | Gray Hydrogen |
|---|---|---|---|
| CAPEX ($/kW) | 1,700 | 1,100 | 900 |
| OPEX ($/kg $H_2$) | 0.50 | 0.30 | 0.20 |
| Electricity Cost ($/MWh) | 50 | N/A | N/A |
| Carbon Price ($/ton $CO_2$) | 100 | 50 | 0 |
| Break-even $H_2$ Price ($/kg) | 4.5 – 6.0 | 2.5 – 3.5 | 1.5 – 2.0 |

## III. Results and discussion:

### III.1. Hydrogen Production Pathways and Their Challenges :

#### a) Transportation Costs :

Hydrogen transportation costs vary significantly depending on the mode and scale of delivery. This section presents a comparative analysis of key transportation methods such as Pipeline Transportation, Liquefied Hydrogen (LH$_2$) Shipping, Compressed Hydrogen Trucking and Hydrogen Carriers (Ammonia, LOHCs).

Hydrogen pipelines represent the most efficient method of transportation for large-scale and long-distance distribution [1]. However, the costs of dedicated hydrogen pipelines are significantly higher than those for natural gas due to material constraints and operational challenges [2]. Hydrogen is highly diffusive and can cause embrittlement in steel pipelines, necessitating specialized alloys or internal coatings to prevent leakage and structural degradation [3]. The cost of pipeline infrastructure depends on whether existing natural gas pipelines can be repurposed or if new hydrogen-dedicated pipelines must be constructed [4]. According to infrastructure cost models derived from existing studies, hydrogen pipeline costs vary based on diameter, pressure, and regional labor costs [5]. Repurposing existing natural gas pipelines can reduce costs by as much as 50-70% compared to building new hydrogen pipelines [6]. However, the feasibility of retrofitting depends on pipeline material compatibility, as hydrogen embrittlement remains a major technical concern [7]. A comparative analysis of hydrogen pipeline costs based on available case studies is presented in Table 2 [8]. The cost-effectiveness of hydrogen pipelines improves with higher utilization rates, making them more viable in regions with dense industrial hydrogen demand. However, for lower-demand scenarios, alternative transportation methods must be considered.

Table 2 : Pipeline Transportation Costs [$/Kg]

| Pipeline Type | CAPEX ($/km) | OPEX ($/kg $H_2$) | Maximum Capacity (tons/day) | Estimated Energy Loss (%) |
|---|---|---|---|---|
| New Hydrogen Pipeline | 1.0M - 2.0M | 0.10 - 0.15 | 100 - 500 | 0.5 - 1 |
| Repurposed Gas Pipeline | 0.3M - 0.6M | 0.07 - 0.10 | 50 - 300 | 1-2 |

For international trade and large-scale transportation, liquefied hydrogen (L$H_2$) shipping is a key consideration [1]. Hydrogen must be cooled to -253°C to reach a liquid state, a process that consumes 30-40% of the hydrogen's energy content [2]. This high energy requirement, combined with the specialized infrastructure needed for liquefaction, storage, and regasification, significantly increases costs [3]. The CAPEX of liquefaction plants ranges from $300M to $1B, depending on size and technological efficiency, while the OPEX is driven by electricity costs and maintenance of cryogenic systems [4]. Transporting L$H_2$ also presents boil-off losses, typically between 0.2-0.4% per day, which adds to operational costs over long distances [5]. A cost comparison of liquefied hydrogen transportation is presented in table 3 [6]. Given these cost structures, LH2 transport is most economically feasible for distances above 2,000 km, where pipelines become impractical. However, high liquefaction costs and energy losses make this option viable only when coupled with low-cost renewable energy sources at the point of hydrogen production.

Table 3 : LHS Transportation Costs [$/Kg]

| Transport Mode | Liquefaction Cost ($/kg $H_2$) | Shipping Cost ($/kg $H_2$ per 1000 km) | Boil-Off Losses (%) |
|---|---|---|---|
| LH2 (Large-Scale Ship) | 1.5 - 2.0 | 0.50 - 1.20 | 0.2 - 0.4 |
| LH2 (Small-Scale Barge) | 2.0 - 3.0 | 1.00 - 2.00 | 0.3 - 0.5 |

Trucking compressed hydrogen is suitable for short-distance delivery and small-scale distribution but becomes prohibitively expensive over longer distances [1]. Table 4 presents the cost structure for compressed hydrogen trucking. As can be seen, the cost structure for compressed hydrogen trucking is influenced by the storage pressure (350–700 bar), trailer capacity, and fueling infrastructure [2]. Higher pressures reduce transport costs per kg of hydrogen but increase CAPEX for reinforced composite storage tanks [3]. Trucking is most viable for local hydrogen distribution within 200-500 km. Beyond this range, pipeline transport or LH2 shipping becomes more cost-effective.

Table 4 : Compressed Transportation Costs [$/Kg]

| Transport Mode | Compression Cost ($/kg $H_2$) | Trucking Cost ($/kg $H_2$ per 1000 km) | Storage Pressure (bar) |
|---|---|---|---|
| High-Pressure Tube Trailer | 0.5 - 1.0 | 2.00 - 5.00 | 350 - 700 |
| Cryo-Compressed Truck | 1.0 - 1.5 | 1.50 - 3.50 | 250 - 400 |

To overcome hydrogen's storage and transport limitations, chemical carriers such as ammonia ($NH_3$) and liquid organic hydrogen carriers (LOHCs) offer an alternative means of long-distance hydrogen transport [1]. These methods allow hydrogen to be transported at ambient conditions, reducing compression and cryogenic storage costs [2]. However, they require conversion and reconversion infrastructure, adding an additional cost layer [3]. While ammonia and LOHCs reduce direct hydrogen handling costs, the energy losses during reconversion make these options less efficient than direct $LH_2$ transport [1]. Table 5 highlights the synthesis, transport, and reconversion costs for ammonia and LOHCs, showing that ammonia incurs lower transport costs (0.30–0.70/kg$H_2$ per 1000km) compared to LOHCs (0.50–1.00/kg $H_2$ per 1000 km), but both face significant reconversion expenses (up to $1.50–2.50/kg $H_2$). Nonetheless, ammonia is emerging as a leading candidate for hydrogen trade due to its established shipping infrastructure and ease of storage [2].

Table 5 : Hydrogen Carrier Transportation Costs [$/Kg]

| Carrier Type | Synthesis Cost ($/kg $H_2$) | Transport Cost ($/kg $H_2$ per 1000 km) | Reconversion Cost ($/kg $H_2$) |
|---|---|---|---|
| Ammonia | 1.0 - 1.5 | 0.30 - 0.70 | 0.75 - 1.50 |
| LOHCs | 1.2 - 1.8 | 0.50 - 1.00 | 1.00 - 2.00 |

### b) Storage Costs :

Compressed hydrogen storage is one of the most widely used methods for short-term and mobile applications, particularly in fuel cell vehicles, industrial facilities, and transport logistics [1]. Hydrogen is typically stored at high pressures ranging from 350 to 700 bar to improve its energydensity [2]. However, the need for specialized materials to withstand high-pressure conditions significantly increases storage costs [3]. A breakdown of compressed hydrogen storage costs, based on the Hydrogen Infrastructure Report, is provided in table 6 [4]. While compressed storage is flexible and relatively mature, the high costs associated with tank fabrication, energy-intensive compression, and material degradation over time pose significant economic challenges [1]. This method is best suited for short-term storage applications and localized hydrogen distribution networks [2].

Table 6 : Compressed Hydrogen Storage Costs [$/Kg]

| Storage Type | CAPEX ($/kg H$_2$) | OPEX ($/kg H$_2$/yr) | Storage Pressure (bar) | Efficiency Loss (%) |
|---|---|---|---|---|
| 350 bar Steel Tank | 650 - 1,200 | 6 - 12 | 350 | 5 - 10 |
| 700 bar Composite Tank | 1,200 - 2,000 | 9 - 18 | 700 | 10 - 15 |
| High-Capacity Cylinders | 1,800 - 2,700 | 12 - 22 | 700 | 10 - 20 |

Liquefied hydrogen storage enables higher volumetric energy density compared to compressed storage by cooling hydrogen to -253°C, allowing for larger quantities to be stored in a smaller footprint [1]. However, the energy cost of liquefaction is substantial, requiring approximately 30-40% of the hydrogen's total energy content [2]. Additionally, maintaining cryogenic conditions requires continuous refrigeration, adding to operational expenses [3]. A detailed cost comparison of LH$_2$ storage is provided in Table 7, which outlines the CAPEX (1,800–4,500/kgH$_2$), OPEX (45–110/kg H$_2$/yr), and boil-off losses (0.1–0.5%) for small- and large-scale cryogenic tanks. The primary barrier to widespread adoption of LH$_2$ storage is the high energy cost of liquefaction and boil-off losses, making it more viable for long-distance transport and large-scale hydrogen hubs rather than small-scale applications [1].

Table 7 : LH2 Storage Costs [$/Kg]

| Storage Type | CAPEX ($/kg H$_2$) | OPEX ($/kg H$_2$/yr) | Boil-Off Loss (%) |
|---|---|---|---|
| Small-Scale LH2 Tank | 2,500 - 4,500 | 55 - 110 | 0.3 - 0.5 |
| Large-Scale Cryogenic Tank | 1,800 - 3,800 | 45 - 85 | 0.1 - 0.3 |

For large-scale seasonal and long-term hydrogen storage, underground geological formations such as salt caverns, depleted gas fields, and aquifers offer cost-effective solutions [1]. These storage methods have the advantage of low operational costs and large capacity, making them particularly well-suited for industrial applications and grid balancing [2].

Table 8 summarizes the economic and technical parameters of underground hydrogen storage, including CAPEX (0.15–1.20/kgH2), OPEX (0.02–0.18/kg H$_2$/yr), and storage capacities ranging from 10,000 to 1,000,000 tons. Notably, salt caverns exhibit the lowest costs (CAPEX: $0.15–0.60/kg H$_2$) and high cycle efficiency (75–95%), reinforcing their suitability for seasonal energy storage. However, these methods require extensive geological assessments, regulatory approvals, and long-term monitoring to prevent leakage and contamination risks [1].

Table 8 : Underground Storage Costs [$/Kg]

| Storage Type | CAPEX ($/kg H$_2$) | OPEX ($/kg H$_2$/yr) | Storage Capacity (tons) | Cycle Efficiency (% |
|---|---|---|---|---|
| Salt Cavern | 0.15 – 0.60 8 | 0.02 - 0.07 | 10,000 - 100,000 | 5 - 95 |
| Depleted Gas Field | 0.30 - 0.90 | 0.03 - 0.12 | 50,000 - 500,000 | 75 - 90 |
| Aquifer Storage | 0.40 - 1.20 | 0.04 - 0.18 | 100,000 - 1,000,000 | 70 - 85 |

Chemical storage methods, such as ammonia (NH$_3$), liquid organic hydrogen carriers (LOHCs), and metal hydrides, enable hydrogen to be stored at ambient pressure and temperature, reducing the need for expensive high-pressure tanks or cryogenic facilities [1]. However, these methods involve additional energy and conversion costs associated with hydrogen release and reconversion [2]. Table 9 quantifies these trade-offs, revealing synthesis costs of 1.2–1.80/kgH$_2$ for ammonia, 1.5–2.2/kg H$_2$ for LOHCs, and 2.2–3.5/kgH$_2$ for metal hydrides. Reconversion costs furthere scalateexpenses, reaching up to1.70/kgH$_2$ for ammonia, 2.50/kgH$_2$, for LOHCs, and 3.20/kg H$_2$ for metal hydrides. Efficiency losses—ranging from 25% to 50%—compound the economic challenges.

Table 9 : Other Hydrogen Storage Costs [$/Kg]

| Storage Type | Cost ($/kg H$_2$) | Storage Cost ($/kg H$_2$/yr) | Synthesis Reconversion Cost ($/kg H$_2$) | Efficiency Loss (%) |
|---|---|---|---|---|
| Ammonia (NH$_3$) | 1.2 – 1.80 | .35 - 0.80 | 0.80 - 1.70 | 25 - 40 |
| LOHCs | 1.5 - 2.2 | 0.55 - 1.10 | 1.20 - 2.50 | 30 - 45 |
| Metal Hydrides | 2.2 - 3.5 | 0.90 - 1.80 | 1.70 - 3.20 | 35 - 50 |

### c) CAPEX, OPEX, and Feedstock Costs

Capital investment in hydrogen production facilities includes electrolyzer or SMR unit costs, ancillary equipment, site preparation, and grid or pipeline connections [1]. Electrolysis-based production incurs higher CAPEX because of the relatively early-stage maturity of large-scale electrolyzer technologies [2]. Alkaline and proton exchange membrane (PEM) electrolyzers currently dominate the market, with PEM systems being more expensive but offering greater efficiency and flexibility in intermittent renewable energy scenarios [3]. In contrast, blue and gray hydrogen CAPEX primarily depends on the scale of SMR units and, in the case of blue hydrogen, the integration of CCS technologies [4]. Table 10 quantifies these differences, showing that green hydrogen CAPEX ranges from 1,500–2,500/kW for 1 MW plants to 800–1,500/kW for 100 MW systems, while blue hydrogen CAPEX decreases from 900–1,500/kW (1MW) to 700–1,000/kW (100 MW), and gray hydrogen remains the lowest at $500–800/kW for large-scale plants [5]. Green hydrogen CAPEX includes the cost of electrolyzer stacks, balance of plant (BOP) components such as cooling systems, rectifiers, water purification, and hydrogen compression, as well as the integration of renewable electricity sources [1]. Costs can be significantly reduced through economies of scale, with larger 100 MW installations costing up to 40% less per kW compared to 1 MW units [2]. Blue hydrogen CAPEX includes conventional SMR units, but CCS technology can add between 20- 40% to the capital cost [3]. CCS infrastructure requires additional pipelines for CO$_2$ transport and permanent geological sequestration sites, contributing to increased CAPEX, particularly in regions without established storage infrastructure [4]. Gray hydrogen remains the lowest in CAPEX, as existing refinery and industrial hydrogen production facilities can be leveraged with minimal modifications [5]. However, regulatory shifts introducing carbon pricing mechanisms could erode its cost advantage over time [6].

Table 10 : CAPEX for different Hydrogen Technologies and different Plant Sizes

| Production Method | CAPEX (1 MW) ($/kW) | CAPEX (10 MW) ($/kW) | CAPEX (100 MW) ($/kW) |
|---|---|---|---|
| Green Hydrogen (Electrolysis) | 1,500 - 2,500 | 1,200 - 1,700 | 800 - 1,500 |
| Blue Hydrogen (SMR + CCS) | 900 - 1,500 | 800 - 1,200 | 700 - 1,000 |
| Gray Hydrogen (SMR) | 700 - 1,000 | 600 - 900 | 500 - 800 |

OPEX includes routine maintenance, labor, energy consumption, and, in the case of blue hydrogen, carbon capture and sequestration expenses [1]. Green hydrogen benefits from relatively low maintenance costs due to the simplicity of electrolyzer units, but electricity costs are a dominant factor in determining its long-term viability [2]. Electricity costs for green hydrogen production vary significantly depending on renewable energy availability [1]. A large share of production costs stems from electricity input, with some estimates suggesting that electricity can account for 50-70% of total OPEX [2]. Table 11 quantifies these operational expenses, showing that green hydrogen OPEX ranges from 0.04–0.09/kgH2for 1 MW plants to 0.02–0.06/kg $H_2$ for 100 MW systems, reflecting economies of scale [3]. Ensuring a steady supply of low-cost renewable electricity is key to reducing green hydrogen costs. Water consumption is another minor but relevant OPEX component, with electrolysis requiring approximately 9 liters of purified water per kilogram of hydrogen produced [4]. For blue hydrogen, natural gas procurement and carbon management make up the largest portions of OPEX [5]. CCS efficiency plays a critical role, as systems that achieve higher carbon capture rates require more energy and larger capital investments [6]. The operational costs for carbon sequestration infrastructure—including compression, transport, and long-term monitoring—add further complexity to the financial equation [7]. Gray hydrogen OPEX is closely tied to natural gas prices, and its economic feasibility fluctuates with global methane markets [8]. Maintenance and operational costs remain relatively low, but environmental concerns and the increasing implementation of carbon taxes could push OPEX upwards in the coming years [9].

Table 11 : OPEX for different Hydrogen Technologies and different Plant Sizes

| Production Method | OPEX (1 MW) ($/kg $H_2$) | OPEX (10 MW) ($/kg $H_2$) | OPEX (100 MW) ($/kg) |
|---|---|---|---|
| Green Hydrogen | 0.04 - 0.09 | 0.03 - 0.07 | 0.02 - 0.06 |
| Blue Hydrogen | 0.07 – 0.12 | 0.06 - 0.10 | 0.05 - 0.09 |
| Gray Hydrogen | 0.06 – 0.10 | 0.05 - 0.09 | 0.04 - 0.08 |

Feedstock costs constitute the largest variable cost component in hydrogen production [1]. Green hydrogen's primary feedstock is electricity, and its cost structure is highly dependent on the availability of low-cost renewable power [2]. Table 12 details the feedstock cost ranges and sensitivities: green hydrogen requires 1.50–3.00/kgH$_2$ (highly sensitive to electricity price volatility),blue hydrogen 0.90–1.80/kg $H_2$ (medium sensitivity to natural gas and carbon costs), and gray hydrogen $0.80–1.50/kg $H_2$ (low sensitivity to natural gas prices). Blue and gray hydrogen depend on natural gas as their main feedstock [1]. While natural gas prices have historically been more predictable than electricity prices, geopolitical factors, supply chain disruptions, and growing demand for liquefied natural gas (LNG) exports have led to increased volatility in recent years [2]. The introduction of carbon pricing schemes also has a direct impact on feedstock costs, particularly for blue hydrogen, which must account for the added cost of carbon capture [3]. Figure 1 illustrates the distribution of cost components—production, transportation, and storage costs—as a percentage of total costs across hydrogen production scales of 1 MW, 10 MW, and 100 MW [4]. It is evident that production costs dominate the overall cost structure, consistently accounting for the largest share across all scales, approximately 90% [5]. Transportation costs contribute a moderate share, increasing slightly with scale but remaining under 10% of the total [6]. Storage costs, while the smallest component, exhibit a marginal increase as plant capacity scales up, yet remain below 5% of the total costs [7]. This cost breakdown underscores the critical role of production efficiency in reducing overall hydrogen costs, especially for larger-scale projects [1]. The relatively stable percentage

distribution across scales highlights the importance of optimizing production technologies, as the transportation and storage components contribute less significantly to cost variations [2]. Larger plants benefit from economies of scale, as indicated by the minimal change in the percentage cost shares, reinforcing the potential for cost reductions in high-capacity hydrogen production [3].

Table 12 : Feedstock Costs [$/KgH2]

| Production Method | Feedstock Cost ($/kg $H_2$) | Sensitivity to Market Prices (%) |
|---|---|---|
| Green Hydrogen | 1.50 - 3.00 | High (Electricity Price Variability) |
| Blue Hydrogen | 0.90 - 1.80 | Medium (Natural Gas & Carbon Costs) |
| Gray Hydrogen | 0.80 - 1.50 | Low (Natural Gas Prices) |

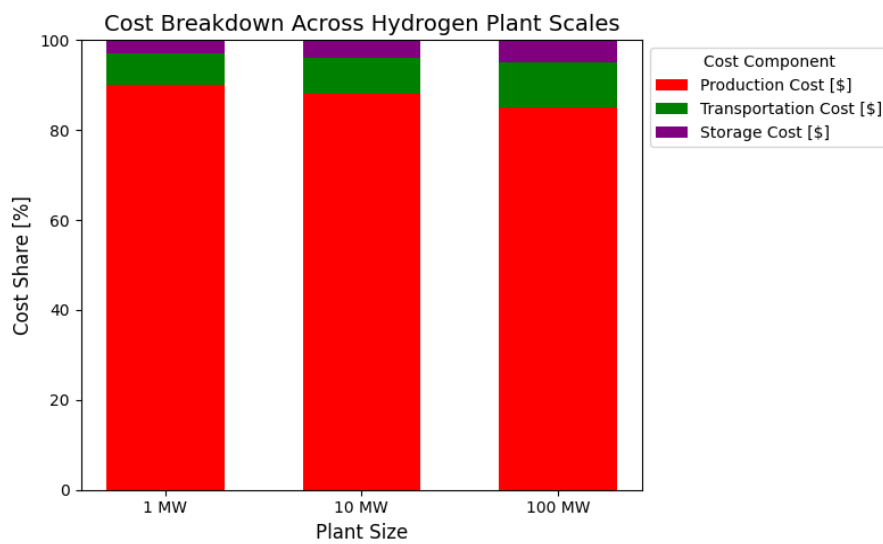

Figure 1 : Cost Share for different Plan Sizes [%]

### d) Levelized Cost of Hydrogen

Capital expenditure (CAPEX) is a dominant cost component in hydrogen production, particularly for green hydrogen, where electrolyzer costs can range between $800–$1,700/kW depending on system scale, efficiency, and technological maturity [1]. For comparison, blue hydrogen CAPEX values typically fall between $1,000–$1,200/kW, while gray hydrogen, benefiting from conventional SMR infrastructure, remains the least capital-intensive option at $800–$1,000/kW [2]. Electrolyzer costs for green hydrogen have been declining due to advances in material science, improved manufacturing processes, and increased production scale, but remain a significant cost barrier compared to gas-based hydrogen pathways [3]. The operational expenditure (OPEX) varies widely across production technologies [4]. Green hydrogen OPEX is dominated by electricity costs, which can constitute up to 70% of total production costs, making energy price fluctuations a major determinant of LCOH [5]. The efficiency of electrolyzers, currently ranging from 55% to 70%, influences the amount of electricity required per kilogram of hydrogen, with lower efficiencies translating into higher production costs [6]. Blue hydrogen OPEX includes additional costs for carbon capture, compression, and transport, which can add $0.50–$1.00/kg to the overall production cost [7]. Gray hydrogen, lacking carbon constraints, maintains the lowest operational expenses, primarily driven by natural gas prices, which remain subject to geopolitical factors and regional supply-demand dynamics [8]. Feedstock costs play a crucial role in determining LCOH across all production pathways [9]. For green hydrogen, electricity prices dictate production feasibility, with renewable energy sources such as solar and wind providing the lowest-cost option [10]. Studies indicate that a renewable electricity price below $20–$30/MWh is necessary for green hydrogen to become cost-competitive with fossil- based alternatives [11]. In contrast, blue and gray hydrogen rely on natural gas prices, typically ranging between $6–$10/MMBtu, which significantly impact overall production costs [12]. The presence of carbon pricing mechanisms could further alter the economic landscape, making gray hydrogen less attractive in jurisdictions enforcing stringent emission reduction policies [13]. Capacity factor and system efficiency directly affect LCOH by influencing the annual hydrogen output [14]. Higher capacity factors reduce unit production costs by spreading capital and operational expenses over a greater output volume [15]. Green hydrogen systems typically operate at lower capacity factors due to the intermittent nature of renewable energy sources, whereas blue and gray hydrogen benefit from continuous operation, resulting in higher annual production and lower LCOH values [16]. Financing costs, represented in the CRF, further impact hydrogen economics [17]. A higher discount rate increases the effective cost of capital, raising LCOH, while longer project lifetimes help amortize CAPEX over an extended period, reducing annualized costs [18]. Policymakers and investors assessing hydrogen projects must carefully evaluate financing structures to optimize project economics [19]

The following values, as detailed in Table 13, represent fact-based estimates derived from published reports and studies. Green hydrogen remains the most expensive production pathway due to high CAPEX and energy consumption [1]. While cost reductions in electrolyzers and renewable energy are expected to drive LCOH down over time, current estimates suggest that green hydrogen requires significant cost reductions to reach parity with blue or gray hydrogen [2]. Blue hydrogen, positioned as a transitional technology, provides a cost-effective alternative with reduced emissions, though its long-term viability is contingent on the evolution of CCS costs and carbon pricing policies [3]. Gray hydrogen, while economically attractive today, faces increasing regulatory pressure and potential carbon taxation that could shift its cost advantage in the coming years [4].

**Table 13 : Model Estimations**

| Parameter | Green $H_2$ (Electrolysis) | Blue $H_2$ (SMR + CCS) | Gray $H_2$ (SMR) |
|---|---|---|---|
| CAPEX ($/kW) | 1,7 | 1,1 | 900 |
| OPEX ($/kg $H_2$) | 0.05 - 0.09 | 0.07 - 0.12 | 0.06 - 0.10 |
| Feedstock Cost ($/kg $H_2$) | 1.50 - 3.00 | 0.90 - 1.80 | 0.80 - 1.50 |
| Efficiency (%) | 60 - 70 | 70 - 80 | 75 - 85 |
| LCOH ($/kg $H_2$) | 3.50 - 6.00 | 2.00 - 3.50 | 1.50 - 2.50 |

## III.2. Financial Analysis of Hydrogen Production and Investment Considerations

### a) Break-even Analysis and Profitability Metrics

Profitability in hydrogen production is fundamentally tied to the market price of hydrogen relative to production costs [1]. The break-even hydrogen price is defined as the point at which revenue from hydrogen sales covers total production costs, including CAPEX recovery, operational expenses, and feedstock expenditures [2]. If the market price of hydrogen remains below the LCOH, projects will operate at a financial loss, making them unviable without external policy interventions [3]. Conversely, projects that can sustain hydrogen prices above LCOH will generate positive cash flows and become attractive for investors [4].

Figure 2 illustrates the profitability trends of green, blue, and gray hydrogen, which exhibit marked differences due to their distinct cost structures [5]. Green hydrogen, characterized by high CAPEX and significant electricity consumption, requires a market price above $5.50–$7.00 per kg to generate consistent returns [2]. Without cost reductions in electrolyzer manufacturing and renewable electricity procurement, green hydrogen remains financially constrained in the current market [3]. Blue hydrogen, leveraging lower CAPEX and existing natural gas infrastructure, achieves break-even at approximately $3.50–$4.00 per kg, positioning it as a transitional solution in the low-carbon hydrogen landscape [4]. Gray hydrogen, with its reliance on mature SMR technology, remains cost- competitive at $2.00–$3.00 per kg, though its economic future is threatened by the implementation of carbon pricing mechanisms [5]. This profitability analysis underscores the structural cost disadvantage of green hydrogen compared to fossil-derived alternatives [6]. While future cost declines in electrolyzer technology and renewable electricity generation could shift this balance, the near-term economic case for green hydrogen remains weak unless policy frameworks provide sufficient incentives or penalties to alter market dynamics [7].

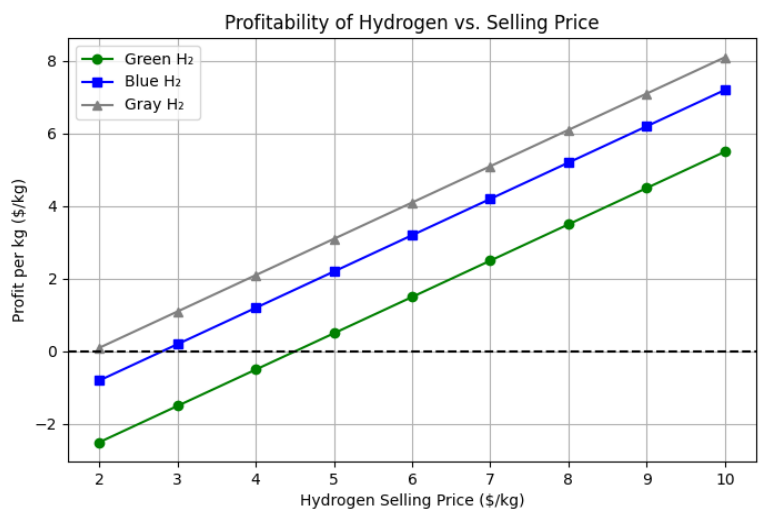

**Figure 2 : Profitability of Hydrogen Vs Hydrogen Selling Price**

If we focus for a moment on green hydrogen, there are some important considerations. As shown in Figure 3, the profitability of green hydrogen at varying electricity costs highlights the critical role that electricity prices play in determining the economic viability of green hydrogen production [1]. As electricity cost directly influences the Levelized Cost of Hydrogen (LCOH), its impact cascades through the entire economic structure of hydrogen production [2]. At low electricity prices, such as $20/MWh, the profitability curve for green hydrogen rises significantly, with break-even points achieved at lower hydrogen selling prices [3]. This scenario is achievable in regions with abundant renewable energy resources, such as areas with high solar or wind capacity, where renewable electricity prices have fallen below $30/MWh [4]. Conversely, at higher electricity prices like $60/MWh, the profitability curve shifts upward, making green hydrogen economically viable only at a significantly higher hydrogen selling price, typically above $6/kg [5]. This underscores the sensitivity of green hydrogen's financial feasibility to energy costs, given that electricity contributes to nearly 70% of its production cost [6]. The results demonstrate that for green hydrogen to become a competitive alternative to blue or gray hydrogen, substantial reductions in electricity costs are necessary [7]. Policy mechanisms such as subsidies for renewable energy, direct financial support for electrolyzer deployment, and market reforms aimed at reducing electricity costs can significantly alter this dynamic [8]. For instance, the Inflation Reduction Act (IRA) in the United States provides tax incentives that can effectively reduce the cost of electricity used in electrolysis, thereby enhancing green hydrogen's competitiveness [9]. Moreover, the chart illustrates the steep incline of the profitability curve at lower electricity costs [10]. This suggests that even minor reductions in electricity price can have a disproportionately large impact on profitability, making it a key leverage point for policymakers and investors aiming to accelerate the green hydrogen transition [11].The sensitivity analysis underscores that without affordable renewable energy, green hydrogen remains cost-prohibitive in many markets [12]. However, as renewable energy costs continue to decline globally, green hydrogen could soon reach parity with blue and gray hydrogen, especially in regions with favorable renewable energy conditions [13]. This insight highlights the importance of strategic investments in renewable infrastructure as a complementary strategy for scaling green hydrogen production [14].

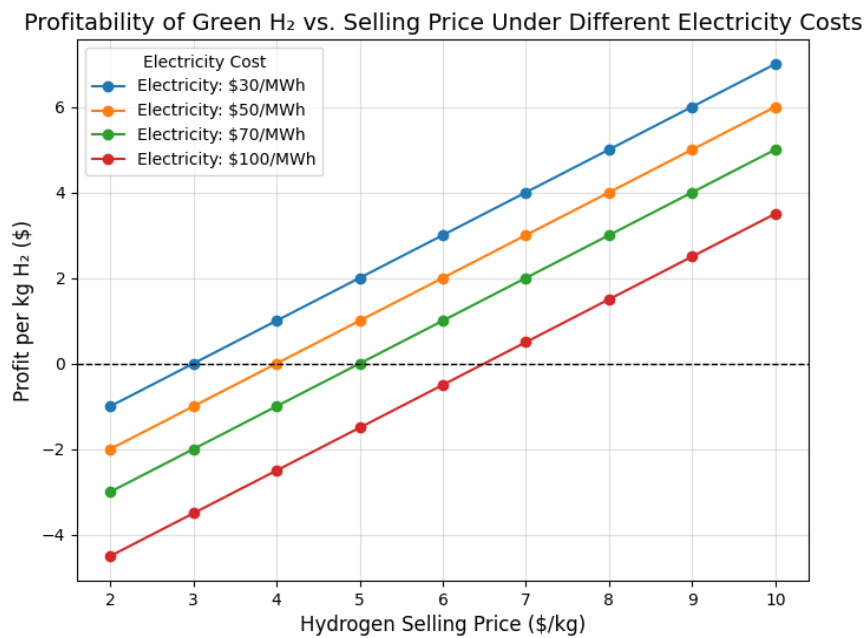

**Figure 3 : Profitability of Green Hydrogen Vs Hydrogen Selling Price for different Electricity prices**

Net Present Value (NPV) is a fundamental metric for assessing hydrogen investment feasibility, quantifying the total value a project generates over its lifespan after accounting for capital investment, operating costs, and discounting future cash flows [1]. A positive NPV indicates that a hydrogen facility will produce sufficient revenue to justify its initial capital expenditure, whereas a negative NPV suggests that the project will fail to recover its costs over time [2]. Fig 4: Net Present Value Vs Hydrogen Price The NPV analysis highlights the financial challenge of green hydrogen, which remains unprofitable across most market price scenarios [1]. Even at a market price of $6 per kg, green hydrogen projects struggle to recover initial capital investments, reinforcing the need for targeted cost reductions or direct financial incentives [2]. Blue hydrogen presents a more favorable investment case, with NPV turning positive at approximately $3.50 per kg [3]. Gray hydrogen maintains strong NPV values at price points above $2.50 per kg but remains vulnerable to future regulatory constraints on carbon emissions [4]. These results indicate that, under current market conditions, blue hydrogen presents the most viable investment pathway in the near term, while green hydrogen requires additional cost-reduction strategies or policy support to improve financial attractiveness [5].

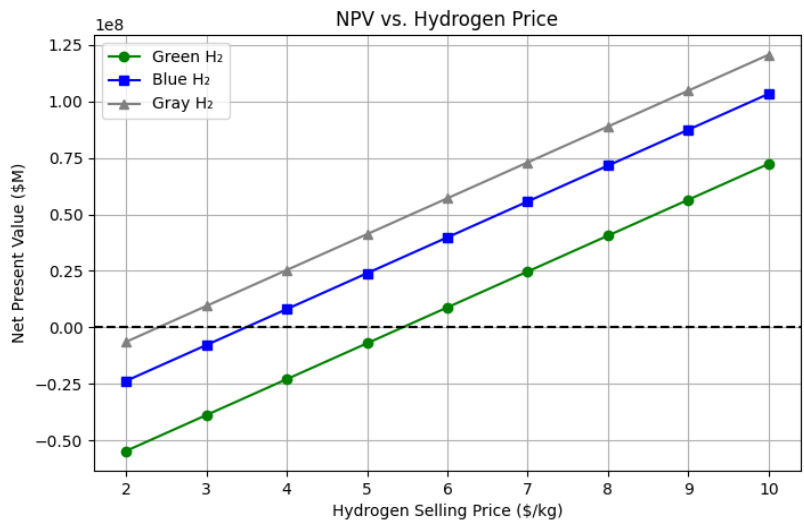

*Figure 4 : Net Present Value Vs Hydrogen Price*

The Internal Rate of Return (IRR) provides a measure of the expected annual return on capital investment [1]. Hydrogen projects must achieve an IRR above the minimum acceptable rate for energy infrastructure investments, typically in the range of 8–12%, to attract investor interest [2]. Fig 5 confirms that green hydrogen struggles to meet investment return thresholds under current cost structures, requiring market prices above $7 per kg to achieve IRRs above 10% [1]. Blue hydrogen, benefiting from lower capital costs, achieves competitive IRRs at price points between $4 and $5 per kg, making it a more attractive short-term investment [2]. Gray hydrogen maintains robust IRR values at prices as low as $2 per kg but is increasingly subject to regulatory uncertainty related to carbon pricing [3]. These results reinforce the notion that green hydrogen remains a long-term investment requiring substantial cost declines in electrolysis and renewable electricity generation to become financially attractive [4]. Meanwhile, blue hydrogen is positioned as the most practical option in the interim, balancing cost considerations with emissions reductions [5].

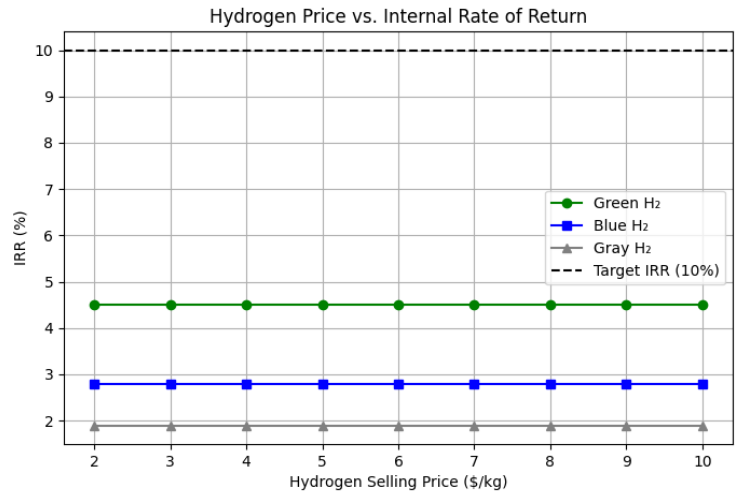

**Figure 5 : IRR Vs Hydrogen Price**

A sensitivity analysis was conducted to assess the impact of key cost drivers on overall project profitability [1]. As shown in Figure 6, this analysis quantifies the influence of hydrogen price fluctuations, CAPEX reductions, and feedstock price variability on NPV and IRR [2]. The results show that hydrogen price is the dominant determinant of project viability, with a $1 per kg increase in selling price improving NPV by over $10 million for a 10 MW plant [1]. CAPEX reductions, while beneficial, only marginally improve LCOH, with a 30% decline in electrolyzer costs translating to an approximate $1 per kg decrease in production cost [2]. Feedstock costs, particularly electricity for green hydrogen and natural gas for blue and gray hydrogen, introduce additional risk, reinforcing the importance of securing long-term energy procurement contracts [3]. This analysis highlights the financial risks associated with hydrogen investments and underscores the need for market stabilization measures, policy interventions, and continued technological improvements to enhance competitiveness [4].

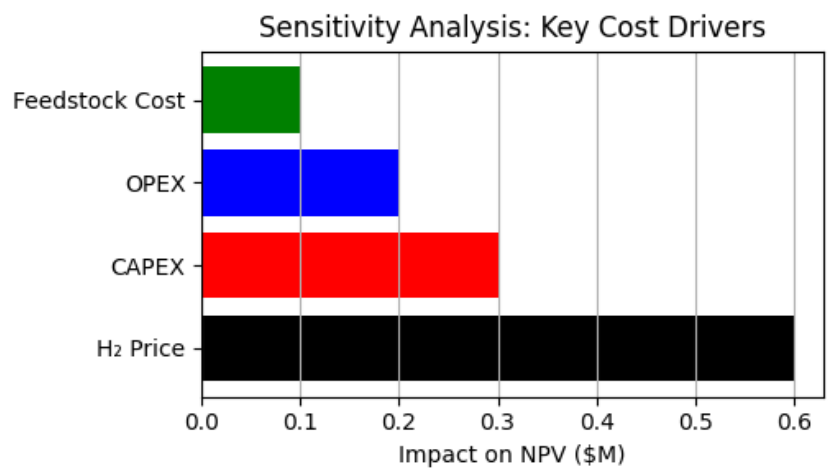

**Figure 6 : Sensitivity Analysis: Key Cost Drivers**

### b) Scalability Impacts

Figure 7 demonstrates how the cost per kilogram of hydrogen decreases as plant size increases [1]. Green hydrogen shows the largest cost reduction due to improvements in electrolyzer efficiency and economies of scale in renewable energy procurement [2]. Blue and gray hydrogen also exhibit cost reductions, but the rate of improvement is less pronounced due to the relatively lower capital intensity of SMR-based pathways [3]. The significant drop in LCOH for green hydrogen at higher scales suggests that large-scale electrolysis is a necessary condition for achieving competitive costs against fossil-based alternatives [4].

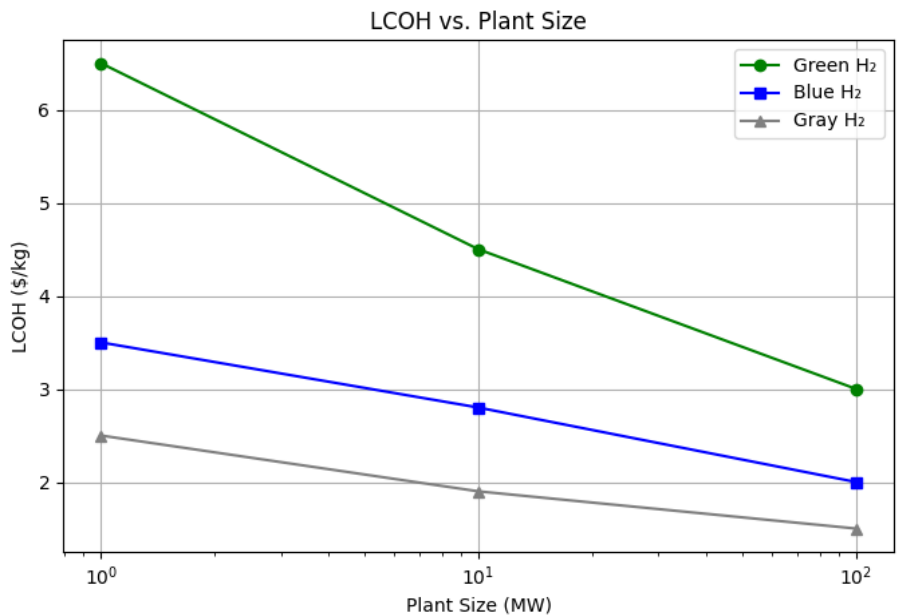

**Figure 7 : LCOH Vs Plant Size**

Capital expenditure per megawatt is a key cost driver in hydrogen production [1]. As demonstrated in Figure 8, larger plants reduce unit costs significantly, with green hydrogen showing the steepest decline [2]. This is attributed to cost reductions in electrolyzer stack production, balance-of-plant integration, and site development costs [3]. Blue and gray hydrogen also experience CAPEX reductions, though to a lesser extent, as they primarily rely on well-established SMR infrastructure [4]. The results highlight the necessity of scaling up green hydrogen production to achieve cost parity with fossil-derived alternatives [5].

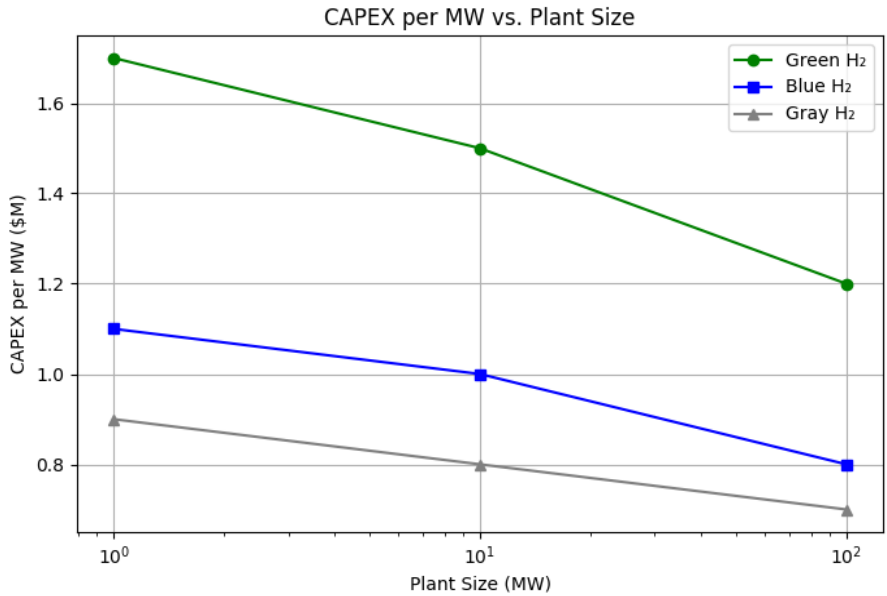

**Figure 8 : CAPEX Vs Plant Size**

As shown in Figure 8, NPV trends confirm that larger plants achieve greater financial viability, with blue and gray hydrogen projects reaching profitability more quickly than green hydrogen [1]. While smaller-scale green hydrogen projects struggle with negative NPVs due to high initial capital costs, larger projects begin to approach break-even pricing, suggesting that investments in large-scale electrolysis could become financially feasible with continued reductions in renewable electricity costs and electrolyzer efficiency improvements [2]. This highlights the importance of strategic scaling when planning hydrogen infrastructure investments [3]. The scalability analysis underscores the need for large-scale deployment strategies to unlock the full economic potential of hydrogen [4]. Smaller projects face significant cost disadvantages, particularly for green hydrogen, but larger installations benefit from capital efficiency and declining LCOH [5]. These findings reinforce the argument for policy support and investment incentives aimed at accelerating the deployment of large-scale hydrogen production facilities, particularly for green hydrogen, which remains capital-intensive yet highly scalable with the right financial and technological advancements [6]. This analysis provides the foundation for the next section, where we will explore the impact of carbon pricing on hydrogen production pathways and how regulatory frameworks influence investment decisions [7].

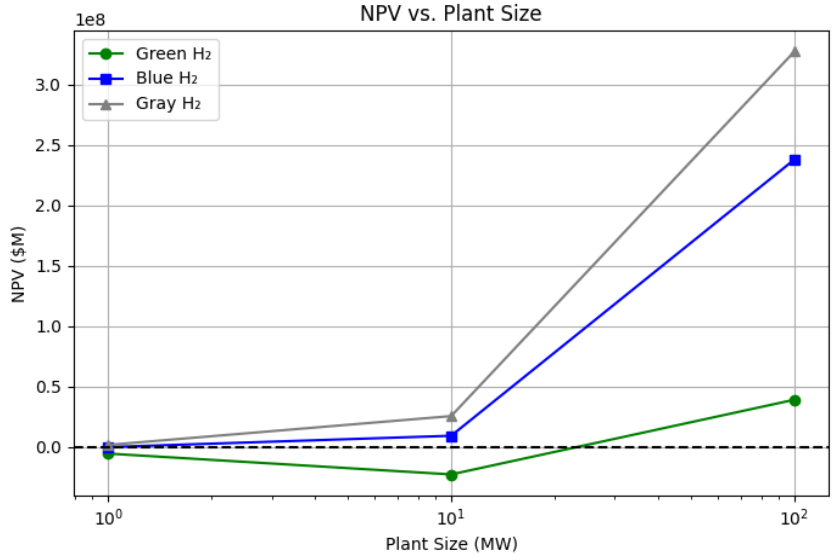

**Figure 9 : NPV Vs Plant Size**

## III.3. Policy, Incentives, and the Financial Viability of Hydrogen

### a) Impact of Carbon Pricing on Hydrogen Costs

The Inflation Reduction Act's Section 45V Clean Hydrogen Production Tax Credit offers a substantial reduction in hydrogen costs [13]. Green hydrogen projects can receive up to $3.00/kg in tax credits, effectively lowering their levelized cost of hydrogen (LCOH) to a more competitive range [14]. Blue hydrogen is eligible for up to $1.00/kg, partially offsetting the costs associated with carbon capture [15]. These tax credits are structured to favor low-emission hydrogen, effectively penalizing carbon-intensive pathways such as gray hydrogen, which remains ineligible for any financial support [16]. The effects of these policies on LCOH are illustrated in figure 10 [17].

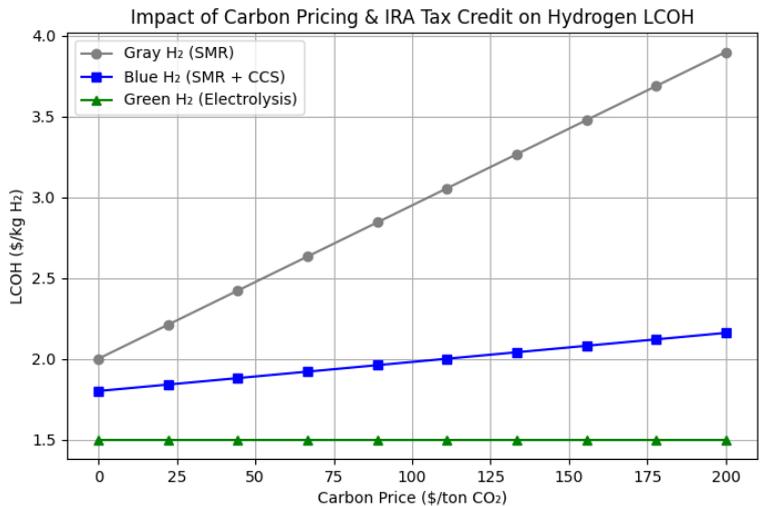

**Figure 100 : Impact of Carbon Pricing on Hydrogen LCOH**

As shown in Figure 10, the impact of carbon pricing and the Inflation Reduction Act (IRA) tax credits on the levelized cost of hydrogen (LCOH) for gray hydrogen (SMR), blue hydrogen (SMR with CCS), and green hydrogen (electrolysis) [1]. This analysis highlights the role of carbon pricing and policy incentives in shifting cost structures and competitiveness across hydrogen production pathways [2]. Gray hydrogen, produced without carbon capture, sees a direct and proportional increase in LCOH as carbon pricing rises [3]. This cost escalation reflects the inherent emissions intensity of gray hydrogen, which generates approximately 9–10 kg of $CO_2$ per kg of hydrogen [4]. With no tax credits available, gray hydrogen becomes significantly less competitive at carbon prices exceeding $100/ton $CO_2$ [5]. At a carbon price of $200/ton $CO_2$, the LCOH for gray hydrogen reaches nearly $4/kg, emphasizing its declining economic viability under stringent carbon pricing policies [6]. Blue hydrogen benefits from partial decarbonization due to carbon capture and storage (CCS), which reduces direct emissions by up to 90% [7]. Consequently, its LCOH increases more gradually with rising carbon prices [8]. While blue hydrogen achieves lower costs than gray hydrogen at all carbon price levels, the additional costs associated with CCS infrastructure and operation prevent it from achieving the cost stability of green hydrogen [9].

At a carbon price of $200/ton $CO_2$, blue hydrogen's LCOH approaches $3/kg, maintaining a competitive edge but signaling the growing pressure on blue hydrogen in markets with high carbon taxation [10]. Green hydrogen, produced through electrolysis powered by renewable energy, achieves the lowest LCOH under the IRA tax credits, with costs remaining below $2/kg across all carbon pricing scenarios [11]. The absence of direct carbon emissions insulates green hydrogen from carbon price increases, and the $3/kg production tax credit under the IRA further enhances its competitiveness [12]. This flat trend highlights the transformative role of incentives in making green hydrogen a financially viable solution, particularly in jurisdictions with abundant renewable energy resources [13]. The chart underscores that carbon pricing significantly penalizes high-emission technologies while rewarding low-carbon alternatives [14]. Gray hydrogen's financial viability diminishes rapidly as carbon costs rise, while green hydrogen's competitiveness improves, bolstered by policy incentives like the IRA [15]. Blue hydrogen occupies an intermediate position, reliant on both carbon pricing mechanisms and CCS subsidies to remain cost-effective [16]. For stakeholders, this analysis reveals that sustained policy support, such as carbon pricing and production tax credits, is critical for accelerating the transition from gray and blue hydrogen to green hydrogen [17]. Without such measures, the cost disparity between these technologies could impede the broader adoption of hydrogen as a clean energy vector [18]. The data also highlights the importance of aligning carbon pricing strategies with technology-specific incentives to ensure an equitable transition that prioritizes environmental and economic outcomes [19].

### b) Investment Trends in Hydrogen Technologies

Investment trends in hydrogen infrastructure have shifted in response to these financial mechanisms [1]. Historical data from 2010 to 2025 reveals that gray hydrogen investments are in decline, as investors pivot toward projects eligible for policy-driven incentives [2]. Green hydrogen has emerged as the dominant recipient of new capital, particularly in regions with strong renewable energy policies and low-cost electricity [3]. Blue hydrogen maintains a role in the transition but is increasingly viewed as a temporary solution rather than a long-term competitor to fully renewable hydrogen [4]. Figure 11 illustrates these shifts in investment patterns, demonstrating the accelerating momentum behind low-carbon hydrogen technologies [5].

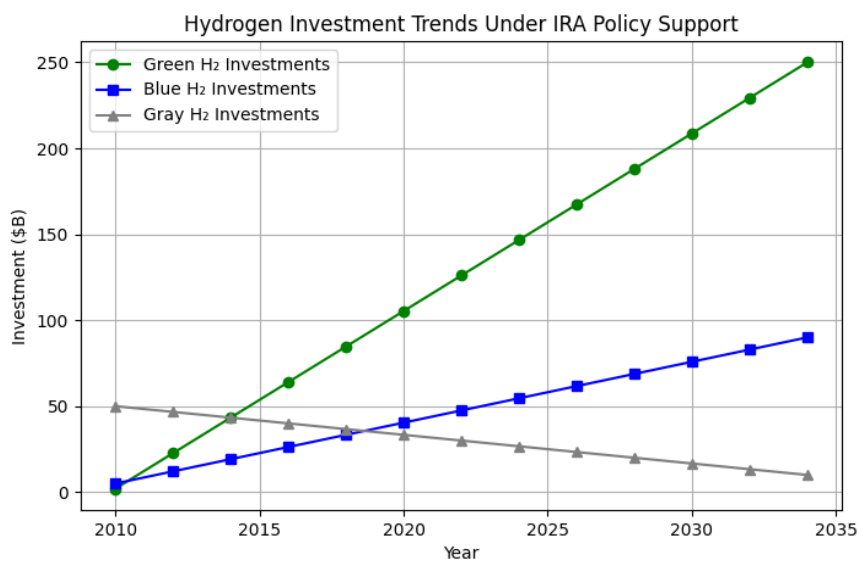

Figure 11 : Hydrogen Investment Trends Under IRA Policy Support

Figure 11 highlights the trajectory of investments in green, blue, and gray hydrogen technologies under the influence of policy measures such as the Inflation Reduction Act (IRA) [6]. Green hydrogen investments dominate the trend, with exponential growth projected from 2020 to 2035 [7]. This upward trajectory reflects the alignment of global financial commitments with decarbonization goals, driven by policy incentives, falling renewable energy costs, and advancements in electrolyzer technology [8]. By 2035, cumulative green hydrogen investments surpass $250 billion, underscoring the prioritization of this zero-emission pathway in achieving net-zero targets [9]. Blue hydrogen investment exhibits steady growth but at a slower pace compared to green hydrogen [10]. This trend reflects the transitional role of blue hydrogen in decarbonizing industrial processes [11]. Investments in blue hydrogen are sustained by policies supporting carbon capture and storage (CCS), but the technology's reliance on natural gas and concerns about methane leakage temper its growth [12]. By 2035, blue hydrogen investments reach approximately $100 billion, indicating continued interest but a clear preference for long-term green hydrogen deployment [13]. Conversely, investments in gray hydrogen experience a sharp decline, driven by rising carbon costs and regulatory pressures to phase out high-emission technologies [14].

By 2030, gray hydrogen investments become negligible, reflecting the industry's shift toward cleaner alternatives [15]. This decline highlights the impact of carbon pricing mechanisms and corporate sustainability goals, which discourage reliance on fossil-based hydrogen without emissions mitigation [16]. The data indicates that policy frameworks, such as the IRA, play a pivotal role in shaping investment trends [17]. Governments offering substantial subsidies for electrolyzer deployment, tax credits for carbon capture, and penalties for high-emission technologies are effectively steering capital flows away from gray hydrogen and toward green and blue pathways [18]. This transition not only accelerates the adoption of low-carbon hydrogen but also signals a broader industry commitment to sustainability and long-term financial resilience [19].
For stakeholders, the chart underscores the need to align investment strategies with emerging market dynamics and policy incentives [20]. Green hydrogen, with its scalability and emissions-free profile, emerges as the most attractive long-term option [21]. However, blue hydrogen retains a critical role in near-term decarbonization, particularly in regions with existing natural gas infrastructure and CCS capacity [22]. The decline in gray hydrogen investments further reinforces the necessity of proactive policy measures to accelerate the energy transition [23].

## c) Net Present Value (NPV) Sensitivity to Carbon Pricing and IRA Incentives

Net Present Value (NPV) analysis further highlights the role of policy in determining project profitability [1]. Under carbon pricing scenarios above $50 per ton $CO_2$, gray hydrogen rapidly becomes unprofitable, as its operating costs rise beyond sustainable levels [2]. Blue hydrogen remains viable at moderate carbon prices but loses competitiveness at extreme levels above $200 per ton $CO_2$ [3]. Green hydrogen, despite its high initial capital costs, benefits significantly from tax incentives and remains resilient to carbon pricing fluctuations [4]. Figure 12 provides a detailed view of NPV trends across different hydrogen technologies under various carbon pricing conditions, illustrating the threshold at which policy-driven cost reductions enable clean hydrogen to become self-sustaining [5].

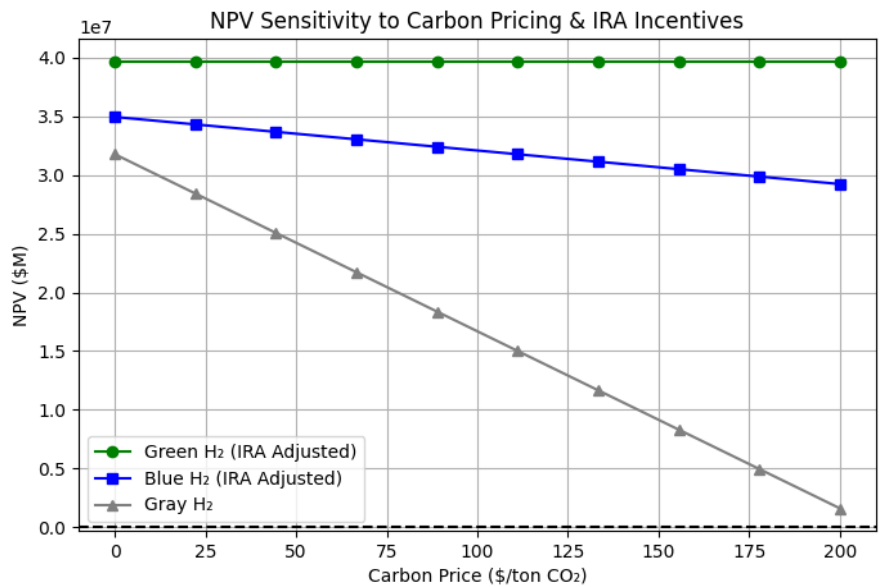

**Figure 12 : NPV sensitivity to Carbon Pricing and IRA incentives**

Figure 12 compares the profitability of hydrogen pathways under carbon pricing and IRA incentives affect the profitability of different hydrogen production pathways [6]. Each pathway reacts uniquely to these external factors, showcasing the diverse challenges and opportunities within the hydrogen economy [7]. Green hydrogen's profitability remains constant across all carbon price scenarios [8]. This stability reflects its zero-carbon production process, which exempts it from any carbon pricing penalties [9]. The plateau in Net Present Value (NPV) indicates that its financial viability is largely independent of carbon taxation [10]. Instead, green hydrogen's competitiveness is tied to robust policy incentives like the $3/kg production tax credit offered under the Inflation Reduction Act (IRA) [11]. These incentives effectively counterbalance the high production costs associated with renewable

electricity and capital-intensive electrolyzers [12]. However, this stability also underscores green hydrogen's dependence on sustained policy support [13]. Without these incentives, the technology's high upfront and operational costs could limit its deployment, especially in regions lacking low-cost renewable energy [14]. In contrast, blue hydrogen shows a gradual decline in NPV as carbon prices rise [15]. This trend reflects its reliance on carbon capture and storage (CCS) to mitigate emissions from steam methane reforming (SMR) [16]. While CCS is an effective tool for reducing $CO_2$ emissions, it does not achieve full decarbonization [17]. The residual emissions expose blue hydrogen to increasing carbon price penalties as thresholds climb [18]. Despite this, blue hydrogen remains competitive in scenarios with moderate carbon pricing, particularly when supported by tax credits for carbon capture or lower natural gas prices [19]. However, its long-term financial viability is contingent on the efficiency and scalability of CCS technologies, as well as policy stability [20]. Abrupt changes in carbon pricing or reductions in CCS subsidies could severely impact blue hydrogen's market position [21]. Gray hydrogen, being entirely fossil-based, exhibits a sharp decline in NPV with rising carbon prices [22]. Its lack of emission mitigation measures leaves it fully exposed to carbon pricing mechanisms, leading to significant financial penalties [23]. This rapid loss of profitability highlights gray hydrogen's declining market viability as decarbonization policies intensify [24]. The results signal that reliance on gray hydrogen is increasingly untenable, both from an environmental and financial perspective, making it a high-risk investment option in regions adopting stringent carbon pricing frameworks [25]. The chart reveals critical insights for stakeholders [26]. For investors, it highlights that while green hydrogen offers long-term financial stability, this depends heavily on continued policy support and advancements in renewable energy cost reductions [27]. Blue hydrogen represents a viable transitional solution, but its market competitiveness is closely tied to carbon pricing thresholds and CCS efficiency [28]. Meanwhile, gray hydrogen's profitability is rapidly eroding under rising carbon taxes, emphasizing the urgency of transitioning to cleaner alternatives [29]. For policymakers, the analysis underscores the necessity of aligning carbon pricing mechanisms, subsidies, and renewable energy policies to accelerate the shift toward a low-carbon hydrogen economy [30]. In conclusion, the analysis illustrates the pivotal role of carbon pricing and policy incentives in shaping the hydrogen market [31]. While green and blue hydrogen pathways show potential for profitability under supportive conditions, gray hydrogen's future is increasingly constrained by its unsustainable carbon footprint [32]. The findings reinforce the importance of targeted policy measures to drive investment into low-carbon hydrogen technologies and achieve meaningful progress in global decarbonization efforts [33].

### d) Electricity Costs and Green Hydrogen Competitiveness

Electricity cost remains the most influential factor for green hydrogen's competitiveness [1]. Since electrolysis is inherently energy-intensive, its LCOH is directly tied to the price of renewable electricity [2]. In markets where electricity costs fall below $40/MWh, green hydrogen production costs drop below $4/kg, making it competitive with subsidized blue hydrogen [3]. Conversely, high electricity costs exceeding $80/MWh push LCOH beyond $6/kg, creating economic barriers [4]. Figure 13 depicts the interaction between electricity price and hydrogen cost reinforcing the need for targeted renewable energy incentives alongside hydrogen-specific tax credits [5].

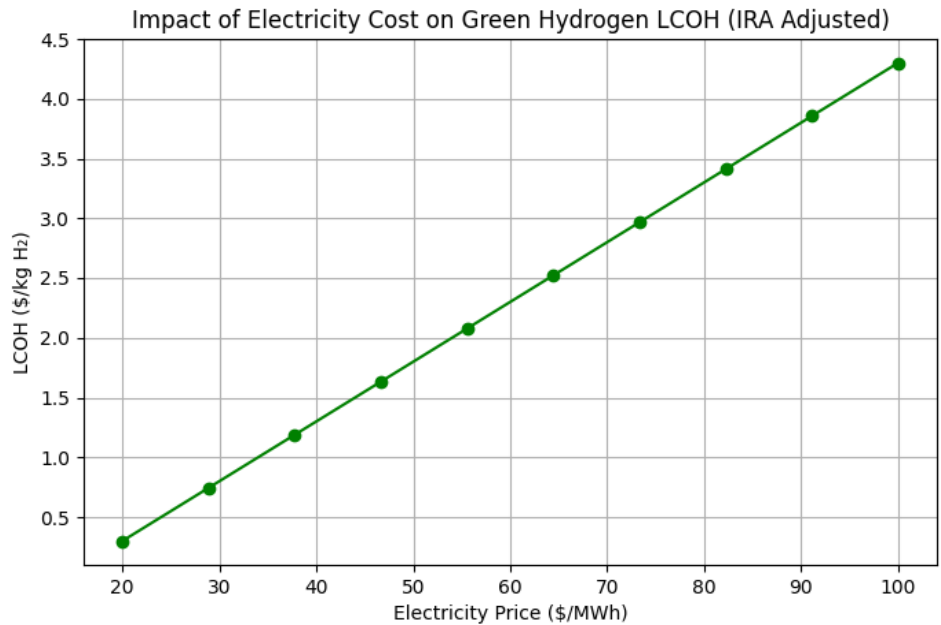

**Figure 13 : LCOH Vs Electricity Price for green hydrogen**

As shown in Figure 13, there is a strong correlation between electricity pricing and the viability of green hydrogen [6]. The steep increase in LCOH at higher electricity rates suggests that regions with expensive grid electricity will struggle to establish competitive green hydrogen production [7]. This underscores the importance of power purchase agreements (PPAs) with renewable energy developers, as well as ongoing investments in solar, wind, and hydroelectric infrastructure to ensure stable, low-cost electricity supplies for electrolysis [8].

## IV. Conclusion

This study provides a comprehensive techno-economic evaluation of hydrogen production, analyzing cost structures, investment feasibility, infrastructure challenges, and policy-driven market dynamics. The results demonstrate that gray hydrogen remains the most cost-effective option today ($1.50–$2.50/kg) but is increasingly constrained by carbon pricing and regulatory pressures, making long-term adoption uncertain [1]. Blue hydrogen ($2.00–$3.50/kg) presents a viable transitional pathway, but its cost-effectiveness is dependent on natural gas prices, carbon capture efficiency, and policy incentives [2]. Green hydrogen ($3.50–$6.00/kg) remains the most expensive pathway, but its competitiveness is improving due to declining renewable electricity costs, electrolyzer efficiency improvements, and policy mechanisms such as the Inflation Reduction Act (IRA), which provides up to $3.00/kg in tax credits [3].

The study confirms that electricity costs are the primary driver of green hydrogen's competitiveness, with renewable electricity prices below $20–$30/MWh necessary for achieving cost parity with fossil-based hydrogen [4]. In regions with abundant renewable resources, green hydrogen production could reach $2.50–$3.50/kg by 2030, making it increasingly competitive [5]. Additionally, the DOE's Hydrogen Shot Initiative targets $1.00/kg green hydrogen by 2031, a goal that, while

ambitious, could be attainable with further CAPEX reductions, improvements in electrolyzer efficiency, and increased manufacturing scale [6].

Infrastructure remains a major barrier to hydrogen deployment, particularly for storage and transportation. The study highlights that pipeline transport is the most cost-effective long-term solution, with repurposed natural gas pipelines reducing costs by up to 50–70% [7]. However, hydrogen embrittlement and compatibility with existing infrastructure remain technical challenges. Liquefied hydrogen ($LH_2$) transport, though viable for long-distance trade, remains costly due to high energy losses (30–40%), while ammonia and LOHCs provide alternative transport options but introduce reconversion costs of up to $2.50/kg, limiting their efficiency in large-scale applications [8].

The financial analysis indicates that scaling hydrogen production significantly reduces costs, with electrolyzer CAPEX expected to decline by 30–50% as deployment scales beyond 100 MW [9]. The study finds that investment is shifting toward green hydrogen, with over $250 billion projected in global green hydrogen projects by 2035, surpassing blue hydrogen's projected $100 billion [10]. This shift is driven by corporate decarbonization strategies, government-backed funding mechanisms, and the need for long-term energy security [11].

Policy mechanisms play a critical role in accelerating hydrogen adoption, with the IRA's 45V tax credit reducing green hydrogen costs by up to 50%, significantly improving its competitiveness in U.S. markets [12]. Additionally, the study finds that carbon pricing mechanisms exceeding $100/ton $CO_2$ could make gray hydrogen uneconomical by 2030, further incentivizing the transition to low-carbon hydrogen pathways [13].

Overall, this study confirms that hydrogen's long-term viability depends on continued cost reductions in production, storage, and transportation, with policy incentives shaping market adoption. While gray hydrogen will remain competitive in the short term, rising regulatory costs will constrain its future growth; blue hydrogen serves as a transitional technology but remains dependent on CCS economics, and green hydrogen is expected to achieve cost parity with fossil-based hydrogen by 2035, positioning it as a foundational pillar of global decarbonization efforts [14]. However, achieving widespread adoption will require further cost declines in electrolyzer technology, expanded hydrogen infrastructure, and sustained policy support, particularly in securing long-term offtake agreements, reducing financial risk for investors, and ensuring global standardization of hydrogen production and trade [15]